\documentclass[11pt,twoside]{article}
\usepackage{atmp04}
\input{epsf.sty}

\copyrightnotice{2003}{7}{121}{144}

\def\gl2{GL(2;\IC)}
\def\nn{\nonumber}
\def\IC{\mathbb{C}}
\def\II{\mathbb{I}}

\def\IR{\mathbb{R}}

\def\IZ{\mathbb{Z}}

\def\IP{\mathbb{P}}

\newcommand{\gen}[1]{\langle #1 \rangle}
\newcommand{\eref}[1]{(\ref{#1})}
\newcommand{\fref}[1]{Figure~\ref{#1}}
\newtheorem{theorem}{\sf THEOREM}

\newcommand{\ba}{\begin{array}}
\newcommand{\ea}{\end{array}}
\newcommand{\bea}{\begin{eqnarray}}
\newcommand{\eea}{\end{eqnarray}}
\def\Hom{{\rm Hom}}

\begin{document}


\title{Closed String Tachyons, Non-Supersymmetric Orbifolds and
Generalised McKay Correspondence}
\arxurl{hep-th/0301162\\}

\author{Yang-Hui He\footnote{
This Research was supported in part by
the gracious patronage of the Dept.~of Physics at the University of
Pennsylvania under cooperative research agreement \# DE-FG02-95ER40893
with the U.~S.~Department of Energy as well as an NSF Focused Research
Grant DMS0139799 for ``The Geometry of Superstrings''. Further
support was kindly offered by the Dept.~of Mathematics of the
Hong Kong University of Science and Technology
as well as the Dept.~of Mathematics
at the Chinese University of Hong Kong under a
``Hodge Theory and Applications in Geometry and Topology'' grant
from the HKRGC.
}}

\address{$^1$Department of Physics,
The University of Pennsylvania,\\
Philadelphia, PA 19104-6396\\
$^2$Department of Mathematics, 
The Chinese University of Hong Kong,\\
Lady Shaw Building, Shatin, Hong Kong.\\
$^3$Department Of Mathematics,
University of Science and Technology,\\
Clear Water Bay, Kowloon, Hong Kong.
}

\addressemail{yanghe@physics.upenn.edu}

\markboth{\it Closed String Tachyons, Non-SUSY Orbifolds and
\ldots}{\it Yang-Hui He}

\begin{abstract}
We study closed string tachyon condensation on
general non-super-symmetric orbifolds of $\IC^2$. Extending previous
analyses on Abelian cases, we present the classification of quotients
by discrete finite subgroups of $\gl2$ as well as the generalised
Hirzebruch-Jung continued fractions associated with the resolution
data. Furthermore, we discuss the intimate connexions with certain
generalised versions of the McKay Correspondence.
\end{abstract}

\newpage
%
\section{Introduction}
Understanding dynamic processes in string theory has recently played
a pivotal r\^{o}le in the field.
Ever since Sen's pioneering work on non-supersymmetric configurations
of
D-branes and their evolution via tachyon condensation
(cf.~e.g. \cite{Sen}),
a host of activities ensued. These tachyonic instabilities,
arising from various scenarios in which supersymmetry is absent, 
shed light on diverse topics ranging from the K-theory
charges lattices to time evolution in cosmology.

Sen's tachyon condensation, which sparked interests such as
the K-theory analysis of D-branes and revival of cubic string field
theory,
focused on the open string sector. There, boundary conformal field
theory
techniques can be applied to track the evolution of the tachyon
toward supersymmetric configurations. The situation in the closed
string
sector is less tractable. We seem to require a full off-shell
formulation
which is thus far not well understood. Indeed it is believed
that the general tachyon condensation process changes the very
structure
of spacetime.

An initial step was taken by Adams, Polchinski and Silverstein (APS)
\cite{panic} where, in analogy to
the provision of defects by D-branes in the open sector, closed
string tachyon condensation was considered on singularities in
spacetime.
By localising the tachyon on so-called non-supersymmetric orbifolds,
more familiar techniques could be used. In the substringy regime,
when the tachyon vacuum expectation value is small, the D-brane probe
technology of \cite{DM} may be applied. 

With this technology we are familiar: we let the D-brane sit not
on flat space, but rather let the transverse dimensions to the brane
be an orbifold singularity. Indeed when the orbifold is Gorenstein
(i.e., we orbifold by a discrete finite subgroup of the special unitary
group),
we have a local Calabi-Yau singularity and some supersymmetry is
preserved. If however we consider subgroups of the general linear
group
as was done in \cite{panic}, supersymmetry is generically broken
and there are tachyons in the tree-level spectrum. 
Therefore in its most prototypical
form and in the notation of \cite{HKMM} we study the propagation of
superstrings on manifolds that could be locally modeled as singular
algebraic varieties of the orbifold type (cf.~\cite{thesis} for
some review on this subject). In other words, the background
geometry of concern is
\begin{equation}
\IR^{d-1,1} \times \IR^{10-d}/\Gamma
\label{compactify}
\end{equation}
where $\Gamma$ is an orbifold group, embedded in some Lie
isometry group of (subspaces of) the $\IR^{10-d}$ factor. The algebraic
structure of $\Gamma$ determines the physics of the transverse
$\IR^{d-1,1}$, the low-energy dynamics of which are of vital
phenomenological importance
\cite{DM,JM,LNV,HanHe,HeSong,Muto,SU4,G2}.

A beautiful
insight of \cite{panic} is that, as is with the open string sector,
the condensation of, i.e., acquisition of VEV's by, the tachyon
leads to a systematic decay of the orbifold that geometrically
corresponds to the partial resolution thereof. The instability is
finally resolved when the decay ends in a supersymmetric
configuration,
viz., when the orbifold finally becomes Calabi-Yau.

Much work followed.
These included renormalisation group (RG) analysis of the
two-dimensional worldsheet field theory \cite{Vafa,HKMM,MM}.
Indeed if we considered the space of 
closed string field theory to be 
the space of two-dimensional field theories then the RG
flow in such spaces will govern the dynamics of strings. 
Therefore we can investigate the
relevant two-dimensional field theories and deformations thereof
corresponding to operators which induce tachyon condensation; 
evolution in physical time can then be identified with the RG flow 
on the worldsheet.  In other words,
tachyon condensation corresponds to the addition of a relevant 
operator to the worldsheet Lagrangian which
describes the background perturbative string propagation. 
The end process is the IR fixed point of the worldsheet RG flow.

Linear sigma model analyses and mirror symmetry
can be applied (to the Abelian cases) 
to track the flow, leading to non-trivial non-supersymmetric 
dualities \cite{Vafa}. Also, a certain $g_{cl}$ conjecture
was proposed in \cite{HKMM}, in identifying the analogue
of the open string boundary entropy which decreases along the flow.
Lifts to M-theory \cite{Uranga,Rabadan} and to F-theory
\cite{Dabholkar}
were also considered as well as addition of flux branes
and Wilson lines \cite{Yi}. Tests of the $g_{cl}$ conjecture
on AdS orbifolds \cite{Basu}, study of bulk condensation 
\cite{Suyama} as well as relations to non-commutative field theory
on $\IC^3$-orbifolds \cite{Uranga2} were performed. 
On a more phenomenological note,
type II and heterotic model-building 
\cite{Font} and chiral phase transitions \cite{Tong} in this context
were also addressed.

In \cite{HKMM,MM} careful analysis was performed on the chiral
ring in the worldsheet conformal field theory and the
tachyon condensation process was geometrised to certain
Hirzebruch-Jung
resolutions of the orbifold $\IC^n / \Gamma \subset GL(n; \IC)$
for Abelian $\Gamma$.

Indeed work thus far has been focusing exclusively on Abelian groups
where methods from toric geometry are happily applicable. A
systematic study using the Inverse Toric Algorithm of \cite{toric}
was performed by \cite{Sarkar} in this context. An obvious direction
beckons us: what about general groups? It is the purpose of this
writing to investigate arbitrary quotients of $\IC^2$ which
do not preserve supersymmetry. We will see that there is a
generalised Hirzebruch-Jung (minimal) resolution for orbifolds
by discrete finite subgroups of $\gl2$ whose classification we present.
We shall also see how extra subtleties arise for non-Abelian
quotients and how they are intimately related to a generalised
version of the McKay Correspondence due to Ishii et al.~and a
conjectured correspondence of \cite{HanHe,HeSong}.

The paper is organised as follows. In Section 2 we define our
problem and briefly review how closed string tachyon condensation
is related to non-supersymmetric orbifolds and how to geometrically
interpret the decay of spacetime. In Section 3 we recast Brieskorn's
classification of the quotients of $\IC^2$ into a form readily
accessible to computation and present a first non-Abelian example 
in the present context. Section 4 then discusses how to view
these issues through generalised McKay Correspondences. Finally we
conclude in Section 5.

%
\subsection*{Nomenclature}
Unless otherwise stated, we shall adhere to the following
notations throughout:
$\Gamma = \gen{a_i | f_j(a_i)}$ is a
discrete finite group of order $|\Gamma|$,
number of conjugacy classes $\#\mbox{conj}(\Gamma)$ and generated
by elements $a_i$ subject to relations $f_j(a_i)$;
The set of irreducible representations of $\Gamma$ is
denoted Irrep$(\Gamma)$, the non-trivial ones,
Irrep$^0(\Gamma)$; The centre of a group $G$ is denoted
$Z(G)$, the derived subgroup of $G$ is written as $G'$,
and $H \triangleleft G$ means that $H$ is a normal subgroup of
$G$;
The primitive $n$-th root of unity is denoted as
$\omega_n := e^{\frac{2\pi i}{n}}$; Finally $(a,b)$ 
means the great common divisor between integers $a$ and $b$.

%
\section{Closed String Tachyon Condensation}
\label{sec:rev}
We shall throughout this writing focus on a subclass of
\eref{compactify}, viz., dimension two orbifolds of the form
\begin{equation}
\label{orb}
\IC^2 / (\Gamma \subset GL(2; \IC) ) \ .
\end{equation}
The localisation by APS of the tachyon to such an orbifold
does not affect the stability of the bulk;
however the local structure of spacetime singularity does change as
the tachyon condenses. APS showed that, as is with Sen's open string case,
the denouement of such a decay process is actually the restoration of
supersymmetry. 

There are two regimes to this description. When the tachyon VEV is small,
we can study the decay at the {\em sub-stringy} scale where one could
apply the usual D-brane probe techniques of \cite{DM}. On the other hand,
when the VEV becomes large, and $\alpha'$ corrections become important,
we are in the {\em gravity} regime and a full worldsheet RG technology
needs to be applied.

We will focus on the brane probe regime. As advertised above, our chief
concern will be two-dimensional quotients. 
In other words, we take $d=6$ in \eref{compactify} and
the orbifold directions transverse to the D-brane
world-volume to be $x^{6,7,8,9}$. We complexify as $z_1 = x^6 + i x^7$ and
$z_2 = x^8 + i x^9$ which constitute the co\"{o}rdinates of $\IC^2$.
In terms of our co\"{o}rdinates, there will be a general twist acting
on $\IC^2$ as $R = \exp(\frac{2 \pi i}{p}( J_{67} + p J_{89} ) )$
with $J_{67}$ and $J_{89}$ being rotations in the $z_1$ and $z_2$
planes respectively.

\begin{figure}
\centerline{\psfig{figure=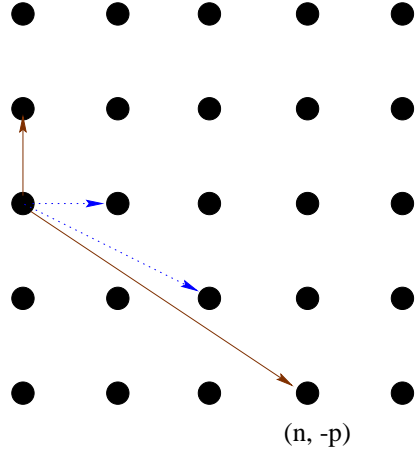,width=4in}}
\caption{The toric diagram for $\IC^2/\IZ_{n(p)}$ consists of single 2D
cone spanned by $v_0 = (0,1)$ and $v_n = (n,-p)$.
Minimal resolution proceeds by inserting lattice vectors $v_j$
in the interior of the cone.
\label{f:Znp}}
\end{figure}

Now we need to consider the general case of \eref{orb}, and
the canonical example is the generalisation of type A, the cyclic
subgroups. Such a quotient, in the notation of \cite{panic,HKMM}, is 
$\IZ_{n(p)}$ (also cf.~\cite{Fulton}) defined as the cyclic group
$\IZ_n$, but with the following matrix action on $(z_1,z_2)$
\begin{eqnarray}
\label{Znp}
\IZ_{n(p)}&=& \gen{\left(
\begin{array}{cc} \omega_n & 0 \\ 0 & \omega_n^p
\end{array} \right)}
\qquad
\omega_n := e^{\frac{2\pi i}{n}}; \\   \quad
n &\in& \IZ^+,
p \in [-(n-1), \ldots, 0, 1, 2,\ldots,n-1], (n,p) = 1 \ .
\nonumber
\end{eqnarray}
Of course $\IZ_{n(-1)}$ is none other than our familiar A-type 
Kleinian singularity (ALE space)
since in this case the group action embeds into $SL(2;\IC)$ 
(see Section \ref{ADE}).
Because of its Abelian nature, the quotient \eref{Znp}
is an affine toric variety
\cite{Fulton} (the reader is encouraged to consult \cite{Sarkar} for
a nice treatment of the toric resolution of such singularities).
The toric diagram is given in \fref{f:Znp}; it consists of a single
two-dimensional cone spanned by $v_0 = (0,1)$ and $v_n = (n,-p)$.

Resolution of toric singularities proceeds by stellar division of the
cone. In other words we insert all vectors between $v_0$ and $v_n$ which
are lattice vectors as shown (in blue)
in \fref{f:Znp}; the resolved space
corresponds to a fan consisting of cones each of which is spanned by
adjacent vectors $v_j$ and $v_{j+1}$. Furthermore there is a relation
\begin{equation}
\label{rel}
a_j v_j = v_{j-1} + v_{j+1} \qquad j \in [1,2,\ldots n-1]
\end{equation}
for non-negative integers $a_j$. Each interior vector $v_j$
then corresponds to an exceptional $\IP^1$ divisor $E_j$, such that
the intersection number are
$E_j \cdot E_{j+1} = 1$ and $E_j \cdot E_j = - a_j$. These integers
$a_j$ are obtained from the so-called {\bf Hirzebruch-Jung}
continued fraction for $\frac{n}{k}$ where $k \in [0,1,\ldots,n-1]$ and
$k \equiv p (\mbox{~mod~} n)$:
\begin{equation}
\frac{n}{k} := a_1 - \frac{1}{a_2 - \frac{1}{a_3 - \ldots}} 
:=
[[a_1, a_2, \ldots, a_r]]
\ .
\label{Jung}
\end{equation}
Indeed as we are expanding a rational, the continued
fraction terminates after finite steps, signifying the
finite number of exceptional divisors. We point out that
this resolution scheme is {\em minimal} in the sense that $r$, the
number of exceptional divisors is smallest amongst all possible
resolutions. In this case all the integers $a_j \ge 2$.

Physically, the exceptional divisors are in one-one correspondence with
the generators of the chiral fusion ring of the ${\cal N}=2$ worldsheet
conformal field theory \cite{Mayr,HKMM}. The chiral
ring structure is captured by the representation ring of the orbifold
group (and hence the quivers which we are about to describe
in Section \ref{classify}), which
in turn is encoded in the intersection numbers of the exceptional
divisors \cite{HeSong,Song}, i.e., with the interior vectors $v_j$.

Now the matter content of the orbifold theory is
simply the McKay quiver associated to $\Gamma$.
We need to be careful that since we here are no longer protected by
supersymmetry we need to construct (generically quite) different
quivers for the fermions and the bosons (cf.~\cite{SU4}).
A quintessential idea of \cite{panic} is that turning on marginal
or tachyonic deformations in the twisted sectors of the orbifold
theory induces partial resolutions of the initial non-SUSY
singularity. The process can be applied consecutively, each
stage being a $\IP^1$-blowup. In other words, the {\em stepwise
tachyonic condensation process corresponds precisely, in the
$\IC^2/\IZ_{n(p)}$ example, to the Hirzebruch-Jung resolution 
outlined in \eref{Jung}.}

As we proceed in this {\em decay of spacetime} where the very topology
of the singular spacetime changes, the symmetry of the theory
subsequently
changes. At each stage of the decay
we should be careful to preserve the quantum
symmetry of the theory while turning on VEV's of tachyons.
Turning on VEV's to break symmetry is of course nothing
but the Higgsing procedure. Indeed, in the case of the supersymmetric
quiver theories, especially the toric ones, such Higgsing procedures
can be systematically studied via the tuning of 
Fayet-Illiopoulos parametres 
and so-called Inverse Algorithm \cite{toric,Sarkar}.

The ${\cal N}=2$ worldsheet CFT provides the link to study the
decay process from a geometric point of view. 
We can parametrise
the $\IC^2$ by the chiral superfields corresponding
to the co\"{o}rdinates. Up to normalisation,
the fusion rules for the chiral ring is dictated precisely by the
Hirzebruch-Jung continued fraction, i.e., by the self-intersection
numbers of the exceptional divisors.
Having translated the problem from chiral rings to geometric
resolutions,
the decay process becomes rather visual. In general $\IC^2/\IZ_{n(p)}$
can decay to an orbifold of lower rank,
\begin{equation}
\label{decay}
\IC^2/\IZ_{n(p)} \rightarrow \IC^2/\IZ_{n'(p')} \oplus \IC^2/\IZ_{n-n'
\ (q)}
\end{equation}
(in particular it can decay to one of
its subgroups in which case $n'$ divides $n$). If the end product is
such
that $p' = -1$ we have then reached our familiar $A_n$ singularity
and hence a supersymmetric orbifold. This is the crucial outcome:
supersymmetry
restoration via tachyon condensation in (localised) closed string
sector.
We remark that the case of $p = +1$ is also, though not immediately
obvious,
supersymmetric because $\IC^2/\IZ_{n(p)}$ and $\IC^2/\IZ_{n(-p)}$ are
isomorphic by conjugation of complex structure: $z_2 \leftrightarrow
z_2^*$.
The resolution data corresponding to \eref{decay}
is of course captured by the respective continued fraction expansions,
the general pattern is
\begin{equation}
\label{decayfrac}
\frac{n}{p} = [[a_1, a_2, \ldots, a_r]] \rightarrow
[[a_1, a_2, \ldots, a_{n'-1}]]
\oplus
[[a_{n'+1}, \ldots a_r]] \ . 
\end{equation}

%
\newpage
\section{The Classification of the Discrete Finite Subgroups of $\gl2$}
\label{classify}
Having refreshed the readers' minds and motivated their hearts
on the interesting physics of
tachyon condensation on the non-supersymmetric orbifold
$\IC^2/\IZ_{n(p)}$, one task is immediate. Can we perform a
similar analysis to all the non-SUSY orbifolds of $\IC^2$? This seems
to require a classification of the discrete finite subgroups of
$\gl2$. At first glance this may appear to be a rather intractable problem
because we seem to have the liberty to quotient $\IC^2$ by any finite
group which affords a two-dimensional irreducible representation,
the list of which is certainly overwhelming. In what follows however
we shall see that it in fact suffices to consider what are known as
{\bf 
small groups} and our candidate pool reduces considerably.
Inspired by the extensive (and still ongoing) programme of the
study of resolutions of quotient singularities, the classification
of $\IC^2$ orbifolds took form in \cite{Brieskorn}.
To a comprehensive presentation in \cite{riemen,Behnke} and especially
in \cite{Behnke} is the reader referred.
Before we proceed however, some preliminary technicalities should be
addressed \cite{Behnke}. To these we now briefly turn.
\subsection{Small Groups and Quotient Singularities}
\label{small}
As mentioned above, we can restrict orbifolds of $\IC^n$ to a very
small subclass. In general we call
a group $\Gamma \subset GL(n;\IC)$ acting on $\IC^n$
a {\bf reflection group} if it is generated by elements
$g$ which fix a hyperplane in $\IC^n$, i.e., the eigenvalues of
$g$ are 1 of multiplicity $n-1$ together with $\omega_{k\ge 2}$, some
root of unity. The complement thereof, i.e., a group which does not
contain any reflections is called a {\bf small group}.

The following Theorems of Prill \cite{Prill} and \cite{Gotts}
therefore narrow down our search considerably and make
the small groups the building blocks of our orbifolds.
\begin{theorem}
\begin{enumerate}
\item
Every quotient singularity is isomorphic to a quotient by a small
group $G \subset GL(n;\IC)$.
\item
If $G_1$ and $G_2$ are small groups in $GL(n;\IC)$ and 
$\IC^n/G_1 \simeq \IC^n/G_2$, then $G_1$ and $G_2$ are conjugate
in $GL(n;\IC)$. 
\end{enumerate}
\end{theorem}
Therefore the classification of conjugacy classes of small groups
of $GL(n;\IC)$ suffices the classification of all complex
quotient singularities.
\subsection{Discrete Finite Subgroups of $SL(2;\IC)$}
\label{ADE}
Before moving on to the small groups, we first set the notation by
briefly reminding the reader that the discrete finite subgroups
of $SL(2;\IC)$ give rise to the so-called Kleinian surface
singularities (cf.~e.g.~\cite{HanHe,thesis} for some implications
in string theory). These groups can be brought, by conjugation,
to the subgroups of $SU(2)$ and fall under 3 types, namely
$A$, $D$ and $E$. Type $A$ is an (reducible)
infinite family which are
abelian, in fact $A_n := \IZ_{n+1}$, the cyclic group on $n+1$
elements. Thus
\begin{equation}
\label{A}
A_{n-1} := \gen{\zeta_n := 
{\scriptsize
\left(
\ba{cc} \omega_n & 0 \\ 0 & \omega_n^{-1} \ea
\right)
}}
\qquad
|A_{n-1}| = n \ .
\end{equation}
Type $D$ is another (reducible imprimitive)
infinite family, the so-called
binary dihedral group, defined as
\begin{equation}
\label{D}
D_n := \gen{\zeta_{2n}, 
\gamma := 
{\scriptsize
\left(
\ba{cc} 0 & i \cr i & 0 \ea
\right)}}, 
\qquad
|D_n| = 4n \ .
\end{equation}
Finally type $E$,
the irreducible primitives, 
consists of 3 exceptional members $E_{6,7,8}$,
respectively the binary tetrahedral, octahedral and icosahedral groups,
generated as
\begin{equation}
\ba{cccc}
E_6 &:=& \gen{S,T}, & |E_6| = 24; \\
E_7 &:=& \gen{S,U}, & |E_7| = 48; \\
E_8 &:=& \gen{S,T,V}, & |E_8| = 120.
\ea
\label{E}
\end{equation}
where
\begin{equation}
\ba{lcl}
S := \frac12 
{\scriptsize
\left(
\ba{cc} -1+i & -1+i \\ 1+i & -1-i \ea
\right)},
&&
T := 
{\scriptsize
\left(\ba{cc}
i & 0 \\ 0 & -i
\ea\right)
},
\\
U := \frac{1}{\sqrt{2}}
{\scriptsize
\left(\ba{cc}
1+i & 0 \\ 0 & 1-i
\ea \right),
}
&&
V := 
{\scriptsize
\left( \ba{cc}
\frac{i}{2} & \frac{(1-\sqrt{5}) - i (1+\sqrt{5})}{4} \\
\frac{-(1-\sqrt{5}) - i (1+\sqrt{5})}{4} & -\frac{i}{2}
\ea \right)
} \ .
\ea
\label{Egens}
\end{equation}

\subsection{Surface Quotient Singularities and Discrete Finite 
Subgroups of $\gl2$}
Now let us return to the $\gl2$ case, the members of which are
generated from the above.
Subsection \ref{small} has shewn us that all relevant
quotients of the type \eref{orb} can be obtained from
the {\em small subgroups} of $\gl2$. We here recast
the classification of \cite{Brieskorn} (cf.~also \cite{Behnke})
into a notation convenient for physical applications.
The interested reader may consult the Appendix on an outline
of how these groups arise.
The small discrete finite subgroups $\Gamma \subset \gl2$
and hence all orbifolds of $\IC^2$ fall under 5 types.
\subsubsection{Type $A_{n,p}$}
\begin{figure}
	\centerline{\psfig{figure=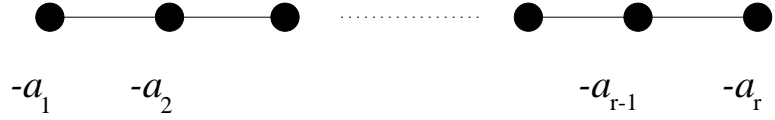,width=4in}}
\caption{The (minimal)
resolution diagram for the type $A_{n,p}$ quotient, i.e.,
$\IC^2/\IZ_{n(p)}$. Each node corresponds to an exceptional
$\IP^1$-divisor, each line corresponds to a single intersection
between two $\IP^1$'s. Each divisor is of self-intersection number
$-a_j$ which is obtained from the continued fraction expansion 
$\frac{n}{p} = [[a_1, a_2, \ldots, a_r]]$.
\label{f:A}}
\end{figure}

This is $\IZ_{n(p)}$, the generalisation 
of the $A_n$ singularity presented in Section \ref{sec:rev}
and studied in \cite{panic,HKMM}. The minimal resolution thereof
is dictated by \eref{Jung}. We can encode the resolution into
a McKay-like quiver \cite{McKay} where each node
corresponds to an exceptional divisor.
The adjacency matrix is given by the
intersection of distinct exceptional divisors. Thus,
the resolution \eref{Jung} is drawn as in \fref{f:A}. 
In the figure, adjacent divisors intersect once while each divisor is
of self-intersection number $-a_j$, obtained from the continued
fraction expansion of $\frac{n}{p}$.
%

\subsubsection{Type $A_mD_n$}
These are composed of the $A$ and $D$ groups; defining
\begin{equation}
\tilde{\zeta_q} :=
{\tiny
\left( \ba{cc}
\omega_q & 0 \cr 0 & \omega_q \ea \right)}
	\subset Z(\gl2) \ ,
\end{equation}
and using the
notation of \eref{A}, \eref{D} and \eref{Egens}, 
there are two subtypes:
\begin{equation}
A_mD_n^{(I)} := \gen{\tilde{\zeta}_{2m}, \zeta_{2n}, \gamma}
\end{equation}
such that $m = (b-1) n - q$ is odd,
as well as
\begin{equation}
A_mD_n^{(II)} := 
\gen{\tilde{\zeta}_{4m}, \zeta_{2n}, \gamma}
\end{equation}
such that $m = (b-1) n - q$ is even.

\begin{figure}
        \centerline{\psfig{figure=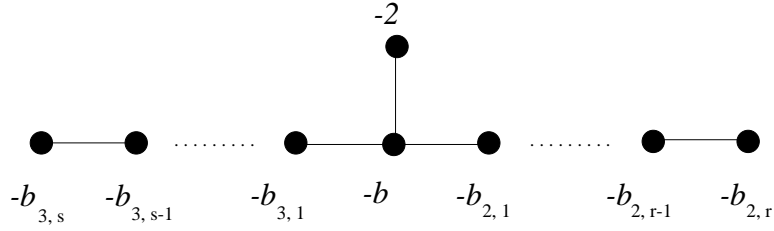,width=4in}}
\caption{The (minimal)
resolution diagram for the general
$\IC^2/\Gamma \subset \gl2$ quotient, to each of which is associated
a determining septuple data ${\cal R} := (b; \quad 2, 1; \quad n_2,
q_2; \quad n_3, q_3)$.
Every node corresponds to an exceptional
$\IP^1$-divisor and every line corresponds to a single intersection
between two $\IP^1$'s. Each divisor is of self-intersection number
$-b$, $-2$,
$-b_{2,j = 1, \ldots r}$ or $-b_{3,j = 1, \ldots s}$, the last two of
which are obtained from the continued fraction of
$\frac{n_i}{q_i} := [[b_{i,1}, b_{i,2}, \ldots b_{i, r(s)}]]$.
\label{f:genreso}}
\end{figure}

The parametres $b$ and $q$ determine the minimal resolution
in the spirit of \eref{Jung}. Let us digress a moment to
settle notation. The adjacency matrix and self-intersection
numbers of the exceptional $\IP^1$ divisors in the
minimal resolution of the general quotient of $\IC^2$
can be associated with the septuple
\begin{equation}
\label{sept}
{\cal R} := (b; \quad 2, 1; \quad n_2, q_2; \quad n_3, q_3)
\end{equation}
or equivalently, with the McKay-type quiver drawn
in \fref{f:genreso}. Again, the nodes correspond to the exceptional
$\IP^1$-divisors and each line signifies the one-time intersection
between two divisors. The self-intersection numbers of each $\IP^1$
are also marked.

As in complete analogy with \eref{Jung} as appeared in 
\cite{HKMM}, we have
continued fractions
\begin{equation}
\label{genJung}
\frac{n_i}{q_i} :=
b_{i,1} - \frac{1}{b_{i,2} - \frac{1}{b_{i,3} - \ldots}} \ ,
\qquad
i = 2,3 \ , 
\end{equation}
such that the intersection of different blowups are given
by the graph in \fref{f:genreso} and the self-intersection
numbers are $-b_{i,j}$ such that when $i=2$, $j$ indexes up to say, $r$
and when $i=3$, $j$ goes up to $s$. Therefore, 
in all the minimal resolution requires $2 + r + s$ exceptional
divisors.

The curious reader may note
that $(2, n_2, n_3)$ always forms a {\em Platonic triple},
i.e., $\frac12 + \frac{1}{n_2} + \frac{1}{n_3} > 1$
and $2 \le n_2 \le n_3$. Moreover, these $A_mD_n$ groups
are two dimensional analogues of what was called ZD-groups
in the brane box constructions of \cite{ZD}.

Bearing this notation in mind, \hfil $A_mD_n^{(I)}$ \hfil  has the
resolution data \newline$(b; \quad 2, 1; \quad 2, 1; \quad n, q)$ and
so too does $A_mD_n^{(II)}$. Henceforth, we shall present
this resolution data ${\cal R}$
in addition to the group structure. Therefore by subtype
of groups we are being a little more refined than
merely referring to the group, but also to its resolution
information.

\subsubsection{Type $A_mE_6$}
There are three subtypes of this category, composed of the cyclic with
the binary tetrahedral:
\bea
A_mE_6^{(I)} &:=& \gen{\tilde{\zeta}_{2m},S,T} \qquad
{\cal R} = (b; \quad 2, 1; \quad 3, 2; \quad 3, 2),
\qquad m = 6(b-2)+1; \nn \\
A_mE_6^{(II)} &:=& \gen{\tilde{\zeta}_{2m},S,T} \qquad
{\cal R} = (b; \quad 2, 1; \quad 3, 1; \quad 3, 1),
\qquad
m = 6(b-2)+5; \nn \\
A_mE_6^{(III)} &:=& \gen{\tilde{\zeta}_{6m},S,T} \qquad
{\cal R} = (b; \quad 2, 1; \quad 3, 1; \quad 3, 2),
\qquad m = 6(b-2)+3 \ . \qquad
\eea
\subsubsection{Type $A_mE_7$}
Next we compose with the octahedral group. There are
4 subtypes. These all have the same structure
\begin{equation}
A_mE_7 := \gen{\tilde{\zeta}_{2m},S,U} ,
\end{equation}
but for different values of $m$, the resolution data differ:
\begin{equation}
\ba{llll}
{\cal R}^{(I)} &=& (b; \quad 2, 1; \quad 3, 2; \quad 4, 3),
	& m = 12(b-2) + 1 \nonumber \\
{\cal R}^{(II)} &=& (b; \quad 2, 1; \quad 3, 1; \quad 4, 3),
	& m = 12(b-2) + 5 \nonumber \\
{\cal R}^{(III)} &=& (b; \quad 2, 1; \quad 3, 2; \quad 4, 1),
	& m = 12(b-2) + 7 \nonumber \\
{\cal R}^{(IV)} &=& (b; \quad 2, 1; \quad 3, 1; \quad 4, 1),
	& m = 12(b-2) + 11 \ .
\ea
\end{equation}
\subsubsection{Type $A_mE_8$}
Finally we compose with the icosahedral group. There are 8 subtypes:
the group is:
\begin{equation}
A_mE_8 := \gen{\tilde{\zeta}_{2m},S,T,V} ;
\end{equation}
again for different values of $m$, the subtypes have distinct
resolution data.
\begin{equation}
\ba{llll}
{\cal R}^{(I)} &=& (b; \quad 2, 1; \quad 3, 2; \quad 5, 4),
	& m = 30(b-2) + 1 \nonumber \\
{\cal R}^{(II)} &=& (b; \quad 2, 1; \quad 3, 2; \quad 5, 3),
	& m = 30(b-2) + 7 \nonumber \\
{\cal R}^{(III)} &=& (b; \quad 2, 1; \quad 3, 1; \quad 5, 4),
	& m = 30(b-2) + 11 \nonumber \\
{\cal R}^{(IV)} &=& (b; \quad 2, 1; \quad 3, 2; \quad 5, 2),
	& m = 30(b-2) + 13 \nonumber \\
{\cal R}^{(V)} &=& (b; \quad 2, 1; \quad 3, 1; \quad 5, 3),
	& m = 30(b-2) + 17 \nonumber \\
{\cal R}^{(VI)} &=& (b; \quad 2, 1; \quad 3, 2; \quad 5, 1),
	& m = 30(b-2) + 19 \nonumber \\
{\cal R}^{(VII)} &=& (b; \quad 2, 1; \quad 3, 1; \quad 5, 2),
	& m = 30(b-2) + 23 \nonumber \\
{\cal R}^{(VIII)} &=& (b; \quad 2, 1; \quad 3, 1; \quad 5, 1),
	& m = 30(b-2) + 29 \ .
\ea
\end{equation}

We thus conclude the presentation of the quotient singularities
of $\IC^2$, i.e., conjugacy classes of the
discrete finite small subgroups of $\gl2$, together with
their quivers for the minimal resolution in the sense of
\fref{f:genreso}, each of which has a generalised
Hirzebruch-Jung continued fraction associated therewith.
We see that all these groups are very simple in that they afford the
direct product structure or a simple quotient thereof. 
Indeed for types $A_nE_{7,8}$ for example, these are direct products
whenever $n$ is odd.
The remarkable fact is that every orbifold of
$\IC^2$ is isomorphic to a member of this simple list above.
%
%
\subsection{A non-Abelian Example}
Armed with this list, let us discuss tachyon condensation associated
therewith.
Thus far all the examples addressed in the literature have been
focusing
on Abelian non-SUSY orbifolds to which toric resolutions in the manner
of \eref{Jung} have been applied. From the classification above, we
see
that the continued-fraction scheme of resolutions
is not limited to the Abelian case
of
type $A_{n,p}$ but persists, via a resolution graph, to all subgroups.

\begin{figure}
        \centerline{\psfig{figure=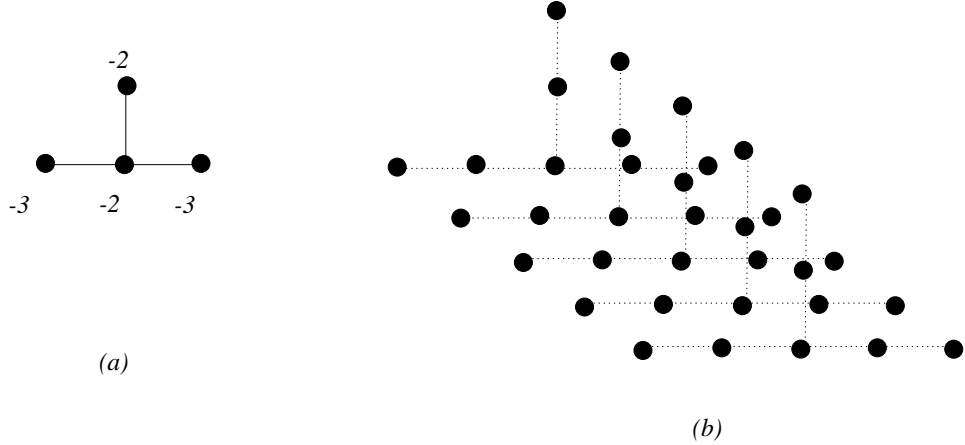,width=5in}}
\caption{(a) The (minimal) resolution graph 
for $\IC^2/A_5E_6^{(II)}$. There are 4 
exceptional divisors.
(b) The McKay quiver for $\IC^2/A_5E_6^{(II)}$, with the fundamental
defining 2-dimensional irrep chosen. As group is simply $E_6 \times
\IZ_5$,
we have 5 copies of 
the familiar $E_6$ quiver interlaced. The dotted lines stand for
how these are linked together, the precise details of which are
not important here. What we wish to emphasize is that
(a) differs markedly from (b) and is only a small subgraph thereof.
\label{f:A5E6}}
\end{figure}

The mathematics of the situation is therefore clear. Let us take the
example of the group $G = A_5E_6^{(II)}$. This is the group
\begin{equation}
G := \gen{\tilde{\zeta}_{10},S,T} \ ,
\end{equation}
which is none other than $E_6 \times \IZ_5$ (where we have chosen
$m=5$ and
hence $b=2$). The resolution data is:
\begin{equation}
{\cal R} = (2; \quad 2, 1; \quad 3, 1; \quad 3, 1) \ .
\end{equation}
The resolution graph is drawn in part (a) of \fref{f:A5E6}, we see
that the
minimal resolution has only 4 exceptional divisors, of self
intersections
$-2,-2,-3,-3$ (the continued fractions in this case are rather
trivial).

As a contrast we present in part (b) of \fref{f:A5E6}, the true McKay
quiver 
for $G$. This quiver is of course the standard one of \cite{DM,LNV}
(cf.~\cite{thesis} or section 2 of \cite{HanHe}
for a review) and dictates the matter content coming form the
projection of the parent theory on the orbifold. In particular we
choose 
the spacetime orbifold action on the fermions (resp.~bosons) as
embedded in the
$SU(4)$ R-symmetry of the parent ${\cal N}=4$ theory on the D-brane
probe; this is
a complex four-dimensional representation $R^4$ of $\Gamma$ (resp.~the
adjoint six 
dimensional). As we do not have supersymmetry here, the 4 and the 6
need not be
related. For illustrative purposes we choose $R^4 = R^1\oplus R^1
\oplus R^2$
where $R^2$ is the fundamental defining irrep of $\Gamma$ and $R^1$ is
the trivial
1-dimensional irrep. Then
\begin{equation}
\label{tensor}
R^2 \otimes R^i = \bigoplus_j a_{ij} R^j
\end{equation}
where $i,j$ indexes over the irreps of $\Gamma$. The finite graph for
which 
$a_{ij}$ is the adjacency matrix is the McKay Quiver; this
gives the (fermion) matter content $a_{ij}$ which counts the
bi-fundamentals 
(The
trivial $R^1$'s give self-linking arrows to each node and are adjoint
fields).
It is this quiver which we draw in part (b) of 
\fref{f:A5E6}. Indeed, because we
have a direct product, the quiver is nothing but 5 copies of 
the familiar $E_6$ quiver appropriately interlaced.

Whereas the mathematics is clear, the physics on the other hand
becomes very involved. Na\"{\i}vely, we could perform a similar
routine
as in \eref{decayfrac} to study the partial resolution of $G$ to one
of its
subgroups, say $\IZ_{5(-1)}$, and we would have the rather curious
process
\[
[[3,3,2,2]] \rightarrow [[1,1,1,1,1]]
\]
where we end up with a longer sequence than we had stated with. The
reason
for this failure has to do with the
weakness of the McKay Correspondence in 
this situation and will be addressed in Section \ref{sec:mckay}.
Furthermore it is not clear at all to what operators these blowup
modes should correspond.

Indeed, the careful analysis of \cite{HKMM,MM} required the one-one 
correspondence between the basis for the chiral ring of the worldsheet
CFT and the exceptional divisors of the resolution so that the
acquisition
of tachyon VEV's and deformations to the CFT may be completely
re-phrased in terms of the geometric resolution of the orbifold. 
This is true for
Abelian orbifolds and thus far only true in this case 
(cf.~e.g.~\cite{HeSong,Song}). For non-Abelian quotients subtleties
arise
\cite{HanHe,HeSong}. The McKay quivers, in comparison with the fusion
graphs
of the corresponding orbifold CFT, needs artificial truncation, i.e.,
only subsectors of the generators of the chiral ring are in
correspondence
with some of the blowup modes of the orbifold. The identification of
which operators to which divisor would be a very interesting problem to
which our list above provides half of the story.
%
%
\section{Tachyons, Orbifolds and McKay Correspondence}
\label{sec:mckay}
We mentioned above that the subtleties involved in the analysis of
the general non-supersymmetric orbifold are related to the Mckay
Correspondence;
in this section let us see this in some detail.

The traditional McKay Correspondence \cite{McKay} dictates that
for the discrete finite subgroups of $SL(2; \IC)$ (q.v.~Section
\ref{ADE}),
the McKay quiver drawn for the defining 2-dimensional representation
as in \eref{tensor} is precisely the Dynkin diagram for the
(simply laced)  affine Lie groups $\widehat{ADE}$, each corresponding
to
the finite group of the same name. The subsequent work on the
geometrisation
of this correspondence (cf.~\cite{Reid} for an excellent account)
hinges
on the fact that for $\Gamma \subset SL(2; \IC)$, there is a one-one
correspondence between the conjugacy classes of $\Gamma$ and the
exceptional 
divisors in the (minimal,
crepant) resolution $\widehat{\IC^2 / \Gamma}$ of $\IC^2 / \Gamma$:
\begin{equation}
\label{cor}
\mbox{conj}(\Gamma) \sim \widehat{\IC^2 / \Gamma}, \qquad \Gamma
\subset  
SL(2; \IC) \ .
\end{equation}

In particular, \eref{cor} implies that the number of $\IP^1$ blowups
is equal to the number of conjugacy classes of $\Gamma$, which in
turn,
by an elementary theorem on the representation of finite groups,
is equal to the number of irreducible representations of $\Gamma$.
Therefore, the number of nodes in the McKay quiver each corresponds to
a $\IP^1$ and the intersection matrix amongst the $\IP^1$'s is the
adjacency matrix of the quiver. Hence we can re-phrase \eref{cor} as
\begin{equation}
\mbox{irrep}^0(\Gamma) \sim H^*(\widehat{\IC^2 / \Gamma}, \IZ)
\end{equation}
where the cohomology picks up the exceptional divisors which are
$-2$ curves and $\mbox{irrep}^0(\Gamma)$ means the non-trivial
irreducible representations of $\Gamma$. 
The nodes of the quiver are usually labeled by the dimension of the
corresponding irrep, which coincides with the Coxeter number of the
associated Dynkin diagram. In summary, the McKay quiver, the minimal
resolution graph for the exceptional $\IP^1$'s and the affine ADE
Dynkin diagrams are identical for $\Gamma \subset SL(2; \IC)$.
Everything therefore fits nicely
for Calabi-Yau orbifolds in dimension two \cite{HeSong}.

For higher dimensions, the correspondence weakens substantially; though
still shown to hold for Abelian quotients in dimension three
(cf.~\cite{ito}),
the generic
group has been so far intractable. In terms of the physics, 
the nice work of \cite{Takayanagi}
for example, has discussed how in the open sector, the $D\bar{D}$
system
exhibits tachyon condensation on three-dimensional 
Calabi-Yau (i.e., $SL(3; \IC)$)
orbifolds and how boundary CFT modes can be interpreted via the McKay 
correspondence from the derived category point of view.
Here we are dealing with another generalisation,
viz., non-Calabi-Yau orbifolds in dimension two and works by Ishii, 
Riemenschneider et al.~shed some light.
\subsection{A Mismatch}
First of all we can immediately see that there is a mismatch between
the conjugacy classes (irreps) and the exceptional divisors for 
$\Gamma \subset \gl2$. The number of nodes in \fref{f:A} is $r$ which
is 
the number of terms in the continued fraction expansion of
$\frac{n}{p}$;
however,
the number of conjugacy classes of this Abelian group is $n$.
As another example, 
in \fref{f:A5E6}, (a) and (b) differ significantly: the
number of exceptional divisors is 4 while there are 35 conjugacy
classes
(irreps). Indeed one can see that this is the general pattern: instead
of
having \eref{cor}, we now have
\begin{equation}
\label{gl2-cor}
\# \mbox{Irrep}(\Gamma) > \# \mbox{Exceptional Div}\left(
\widehat{\IC^2 / \Gamma} \right), \qquad \Gamma \subset  
GL(2; \IC) \ .
\end{equation}

This discrepancy of course shows up in the physics. The twisted
sector operators in the orbifold CFT are counted by the conjugacy
classes of the group (we can see this when
summing up the torus partition function). When we are dealing with
D-brane probes, the equivariant K-theory of the orbifold gives the
representation ring of $\Gamma$ and so there are still enough distinct
D-branes at the orbifold point. However in the resolution of the
singularity, \eref{gl2-cor} means that there are not enough cycles
for the distinct D-branes to wrap. In a very nicely
detailed gauged linear sigma model analysis, \cite{MM} has shown that
in the toric case (i.e., $\IZ_{n(p)}$), while $r+1$ charges come
from the Hirzebruch-Jung resolution (Higgs branch), 
an extra set of $n-r-1$ D-branes
lives on the Coulomb branch whereby conserving the total D-brane
charges.

We remark that this mismatch phenomenon is quite generic and
stands largely unresolved for arbitrary quotients.
It was pointed out in \cite{HeSong}
that the ADE meta-pattern, ubiquitous in
various fields, seems to enjoy the specialty of Calabi-Yaus in
dimension two. Indeed as was shown in \cite{HanHe}, attempts
to relate the chiral fusion ring and twisted sector operators
associated with blowups (i.e., the full geometrisation
of the orbifold conformal field theory), even in the Calabi-Yau
case for dimension 3 met similar mismatches as \eref{gl2-cor}.
Graphical truncation of the quivers are needed and only subsets
of the chiral operators can be placed in correspondence with
either the representation ring of the group, or with
the exceptional divisors in the resolution.

\subsection{Ishii's McKay Correspondence for $\gl2$}
For the general two-dimensional quotients
we should not be discouraged by the mismatch mentioned
above, and there is a
partial
remedy. For this we need to first turn to the work by Ito and Nakamura.
It is known that the symmetric orbifold, $(\IC^2)^n / S_n$ where $S_n$
is
the symmetric group on $n$ elements, has a smooth resolution, the
so-called
Hilbert Scheme of points on $\IC^2$, denoted as Hilb$^n(\IC^2)$. It
was 
realised in \cite{ito} that if one took $n =
|\Gamma|$ and
$\Gamma \subset SL(2; \IC)$,
and considered the $\Gamma$-equivariant version of the said
resolution, one
in fact has
\begin{equation}
\widehat{\IC^2 / \Gamma} = \mbox{Hilb}^\Gamma(\IC^2) := 
(\mbox{Hilb}^n(\IC^2))_\Gamma \rightarrow 
\left((\IC^2)^n / S_n \right)_\Gamma \simeq \IC^2 / \Gamma \ .
\end{equation}
This construction distinguished the Hilbert scheme as the (crepant)
minimal
resolution and provided an explicit mapping of the (non-trivial)
irreducible 
representations with the exceptional divisors.

Into the Hilbert scheme resolution we shall not delve too far. The key
point
for our purposes is the following theorem due to A.~Ishii:
\begin{theorem}
For small subgroups $\Gamma \subset \gl2$, 
the Ito-Nakamura construction for the
McKay Correspondence still holds if instead of considering
Irrep$^0(\Gamma)$,
we consider a subset, the so-called {\bf special representations}.
\end{theorem}
The definition of these representations is rather technical and
we briefly touch upon it. Let $\mu : \IC^2 \rightarrow \IC^2 / \Gamma$
be the canonical projection and $\pi : \widetilde{\IC^2 / \Gamma}
\rightarrow \IC^2 / \Gamma$ be the resolution for $\Gamma \subset
\gl2$.
Moreover let $V_{\rho^*}$ be the representation vector space (module)
associated to the dual $\rho^*$ of the irrep $\rho$ and 
${\cal O}_{\IC^2}$ the sheaf of holomorphic functions on $\IC^2$. We
can then define (a so-called reflexive sheaf) 
$M_\rho := \mu_* \left( {\cal O}_{\IC^2} \otimes V_{\rho^*}
\right)^\Gamma$.
A representation $\rho$ is {\em special} if
\begin{equation}
\label{special}
H^1(\widetilde{\IC^2 / \Gamma}; (\pi^* M_\rho)^*) = 0 \ .
\end{equation}

The upshot is that we can check \eref{special} against all the
(non-trivial)
irreducible representations - easily read off from the character
tables -
of the groups in the classification of Section
\ref{classify}; the only ones which satisfy the condition will be in
one-one
correspondence with the exceptional divisors in the minimal
resolution,
i.e., with the nodes in the resolution quivers (\fref{f:genreso}) and
with the
terms in the generalised Hirzebruch-Jung continued fraction
\eref{genJung}.
In the Abelian case of type $A_{n,p}$, these special representations
are associated to the $r+1$ ``obvious'' D-brane charges sitting at the
Higgs
branch as discussed in \cite{MM}.
\section{Discussions and Prospects}
In this paper we have generalised the problem of closed
string tachyon condensation of \cite{panic}
from the Abelian cases thus far
addressed in the literature to non-Abelian groups.
In particular we have considered arbitrary quotients of
$\IC^2$. To do so we have presented the classification
of all relevant discrete finite subgroups $\Gamma \subset \gl2$,
namely the so-called small groups, the orbifolds thereby
exhausts all isomorphism classes of non-supersymmetric
orbifolds in dimension 2.

We have initiated the study of identifying tachyonic
operators with blowups in the resolution of the orbifolds by
D-brane probes
and pointed out generalisations of the Hirzebruch-Jung
continued fractions used in the toric analysis of the Abelian cases
\cite{HKMM,Sarkar}. A first non-Abelian example, namely type
$A_5E_6^{(II)}$, has been studied in detail.

This geometrisation of the spacetime decay process is
unlike supersymmetric orbifolds of the ALE spaces 
(local K3's) where the full McKay Correspondence conveniently
gives one-one correspondences amongst the chiral ring generators,
the irreducible representations of $\Gamma$ as well as the
exceptional divisors in the (minimal, crepant) resolution of the
orbifold. Here we generically have far more conjugacy classes
(and thus D-brane types) than geometric cycles and we
turn to Ishii's generalised McKay Correspondence for $\gl2$.

The work of \cite{MM} showed how to identify the ``missing 
cycles'' by going to the Coulomb branch to seek
more D-brane charges from the K-theory. The general problem
of matching the chiral ring, conjugacy class and blowups for
arbitrary orbifolds \cite{HanHe,HeSong} still remains tantalising.
We have provided the geometric data for non-supersymmetric
non-Abelian quotients of $\IC^2$, and it would be interesting to
extend the analysis of \cite{MM} in finding the missing
branes. Such a search would be guided by \eref{special}. The ``obvious
branes charges'' associated to the fractional branes that wrap cycles
which by Ishii's correspondence are in 1-1 relation with the special
representations. 

In this light if we were to use a world-sheet CFT
analysis to find the extra charges we could find a strong version of
generalised McKay \cite{MM}. This seems to be a {\em general philosophy:}
even though there is a bijection between the representation ring of
the orbifold group $\Gamma$ and the exceptional divisors of the geometric
orbifold $\IC^n$ known only for the classical case of $\IC^2/ \{
\Gamma \subset SU(2) \}$, string theory knows more. By the very virtue
that the geometric resolution of the orbifold occurs only for the Higgs
branch of the world-volume theory, the other phases of the
moduli space should provide additional information. Could this 
help establish bijections between Rep($\Gamma$) and
$H^*(\IC^n/ \Gamma)$ and in particular 
explain the graph truncations of \cite{DiFrancesco,HeSong}?

Mathematically\footnote{We thank Prof.~Y.~Ruan for
	discussions on this.}
this raises another fascinating point. In dimensions
great than or equal to 2, the aforementioned
mismatch between the blowups
and conjugacy classes of the orbifold for the generic orbifold
is only part of the problem: in general there is not even
a notion of minimal resolution. If we could generalise the linear
sigma model technique in studying the world-sheet CFT
and identifying twisted operators with the blowups, could we
physically distinguish a resolution of the orbifold
(in a sense analogous
to the Hilbert scheme resolution of Ito-Nakamura being a
distinguished one)?

Moreover, it is well known that the ALE spaces
do not admit discrete torsion in the sense that for
$\Gamma \subset SU(2)$, the Schur multiplier
$M(G) := H^2(\Gamma; U(1)) = 0$ (cf.\cite{tor}).
However, now due to
the direct product structure, we can have
non-trivial
projective representations of the D-brane probe.
In general we know that (see e.g.\cite{tor})
\begin{equation}
M(G_1 \times G_2) \simeq M(G_1) \times M(G_2) \times
\Hom_\IZ(G_1/G_1', G_2/G_2') \ ,
\end{equation}
which for those members which afford
the direct product structure, 
e.g.~the cases of the for $A_nE_{7,8}$ ($n$ odd),
simplify to $\Hom_\IZ(G_1/G_1', G_2/G_2')$ because
each $G_{i=1,2}$ is an $SL(2;\IC)$ group and has trivial Schur
multiplier. Also we know that $A_n / A_n' \simeq A_n$ 
because $A_n$ is
Abelian. Moreover, we know that $D_n/D_n' \simeq \IZ_4$,
$E_6 / E_6' \simeq \IZ_3$,
$E_7 / E_7' \simeq \IZ_2$, and $E_8 / E_8' \simeq \II$. Finally,
recalling that $\Hom_\IZ(\IZ_m,\IZ_n) \simeq \IZ_{(m,n)}$, we can
easily determine $M(G)$ for these product groups. Indeed
we see that
only the orbifolds of type $AD$ and $AE_{6,7}$ admit discrete torsion.
For example the group $A_9E_6^{(III)}$ has a $\IZ_3$
discrete torsion. Therefore there should 3 disconnected pieces of the
quiver diagram if one turns on the NS-NS B-field, 
one for each value of the torsion.

It would be worthwhile to investigate
these torsion examples, the moduli space of the gauge theory
as well as possible notions of non-commutativity and
local mirror symmetry in such situations of dimension two. 
These and many other
intriguing issues that stem from these non-supersymmetric orbifolds
we leave to future investigation.

%

\section{Appendix: From $SL(2;\IC)$ to $\gl2$}
We can obtain all the isoclasses of the discrete finite subgroups
of $\gl2$ from the ADE groups of $SL(2;\IC)$. We first note that
there is a surjective homomorphism
\bea
\psi &:& Z \times SL(2;\IC) \rightarrow \gl2 \nonumber \\
&&	(z, g) \rightarrow zg
\eea
where $Z$ is the centre of $\gl2$. Because $Z$ commutes
with all elements, we can generate the appropriate
subgroup $\Gamma$ by concatenating the
generators of the subgroup of $Z$ with that of $SL(2;\IC)$.

Subsequently, all the non-cyclic (the cyclic ones are
the $\IZ_{n(p)}$) finite subgroups $\Gamma$ of $\gl2$ may be
obtained from the quadruple
\[
\Gamma := (G_1, N_1; G_2, N_2)
\]
where $N_1 \triangleleft G_1 \subset Z$ and
$N_2 \triangleleft G_2 \subset SL(2;\IC)$ with $G_2$ not cyclic.
Furthermore, the factor groups are isomorphic as
\[
\phi : G_1 / N_1 \stackrel{\sim}{\rightarrow} G_2 / N_2 \ .
\]
Under these conditions, we can construct $\Gamma$ explicitly
as
\[
\Gamma = \psi(G_1 \times_{\phi} G_2)
\]
where $G_1 \times_{\phi} G_2 := \left\{
(g_1,g_2) \in (G_1,G_2) | \bar{g_2} = \phi(\bar{g_2})
\right\}$ with $\bar{g_i}$ being the residue class 
associate to $g_i$ under the
quotient $G_i/N_i$. In fact the conjugacy class of $\Gamma
\subset \gl2$ is independent of the chosen isomorphism
$\phi$ and so the quadruple suffices to denote the group.

\section*{Acknowledgements}
{\it Ad Catharinae Sanctae Alexandriae et Ad Majorem Dei Gloriam...\\}
I am much indebted to Professors 
Oswald Riemenschneider and Yong-Bin Ruan for their invaluable
suggestions and comments. To the gracious patronage of
the Dept.~of physics at the University of Pennsylvania as well
as the warm hospitality of Yong-Bin Ruan of the Hong Kong
University of Science and Technology and Zuo Kang of the Chinese
University of Hong Kong I am most obliged.
To the ever-delightful smile of Ravi Nicolas Balasubramanian
and the over-flowing cup of William ``Buck'' Buchanan I dedicate this
work.

I am grateful to Kazutoshi Ohta for originally interesting me in
this problem. Comments on the first version of the paper by
R.~Britto-Pacumio and A.~Adams are much appreciated.
Furthermore, conversations with
Akira Ishii and the ``Derived Categories Discussion
Group'' during the ``School for Geometry and String Theory'' at the
Isaac Newton Institute for Mathematical Sciences -- wonderfully
organised by the Clay Mathematics
Institute and the University of Cambridge, as well as with Jessie
Shelton, during the most jovial wedding of the good Capt.~Mario Serna
and the lovely Miss Laura
Evans, are thankfully acknowledged.


\end{document}